\documentclass[twocolumn,showpacs,preprintnumbers,amsmath,amssymb,prb]{revtex4}

\usepackage{graphicx}
\usepackage{dcolumn}
\usepackage{bm}

\begin{document}


\title{Ferroquadrupole ordering and $\bold \Gamma_5$ rattling motion\\in clathrate compound Ce$_3$Pd$_{20}$Ge$_6$}

\author{Yuichi Nemoto, Takashi Yamaguchi, Takenobu Horino,\\ Mitsuhiro Akatsu, Tatsuya Yanagisawa, and Terutaka Goto}
\affiliation{Graduate School of Science and Technology, Niigata University, Niigata 950-2181, Japan}

\author{Osamu Suzuki}
\affiliation{National Institute for Materials Science, 3-13 Sakura, Tsukuba 305-0003, Japan}

\author{Andreas D\"{o}nni}
\affiliation{Department of Physics, Niigata University, Niigata 950-2181, Japan}

\author{Takemi Komatsubara}
\affiliation{Center for Low Temperature Science, Tohoku University, Sendai 980-8578, Japan}

\date{\today}

\begin{abstract}
Lattice effects in a cerium based clathrate compound Ce$_3$Pd$_{20}$Ge$_6$ with a cubic Cr$_{23}$C$_6$-type structure 
have been investigated by ultrasonic and thermal expansion measurements. Elastic softenings of $(C_{11}-C_{12})/2$ and 
$C_{44}$ proportional to the reciprocal temperature $1/T$ above $T_{\rm Q1}=1.25$ K are well described 
in terms of the quadrupole susceptibility for the ground state $\Gamma_8$ quartet. 
A huge softening of 50 \% in $(C_{11}-C_{12})/2$ and a spontaneous expansion $\Delta L/L=1.9\times10^{-4}$ 
along the [001] direction in particular indicate the ferroquadrupole ordering of $O_2^0$ below $T_{\rm Q1}$. 
The elastic anomalies associated with the antiferromagnetic ordering at $T_{\rm N2}=0.75$ K and 
the incommensurate antiferromagnetic ordering are also found. Notable frequency dependence of $C_{44}$ around 10 K is 
accounted for by the Debye-type dispersion indicating a $\Gamma_5$ rattling motion of an off-center Ce ion along the [111] 
direction with eight fractionally occupied positions around the 4a site in a cage. 
The thermally activated $\Gamma_5$ rattling motion obeying a relaxation time $\tau = \tau_0$exp$(E/k_BT)$ 
with an attempt time $\tau_0=3.1\times10^{-11}$ sec and an activation energy $E=70$ K dies out with decreasing temperature, 
and then the off-center tunneling state of Ce ion in the 4a-site cage will appear at low temperatures.
\end{abstract}

\maketitle

\section{INTRODUCTION}
The 4f-electronic systems with spin and orbital degrees of freedom in rare earth compounds frequently reveal electric 
quadrupole orderings in addition to magnetic dipole orderings at low temperatures. The cubic compounds based on Ce$^{3+}$ 
ion with a $\Gamma_8$ quartet ground state in particular have received much attention because the competitive phenomena 
associated with magnetic dipole, electric quadrupole and magnetic octupole degrees of freedom are expected. 
The direct product of $\Gamma_8 \otimes \Gamma_8$ is reduced to a direct sum 
$\Gamma_1 \oplus \Gamma_2 \oplus \Gamma_3 \oplus 2\Gamma_4 \oplus 2\Gamma_5$. 
The magnetic dipole $J_x$, $J_y$, $J_z$ belonging to $\Gamma_4$ symmetry are order parameters for magnetic orderings. 
The quadrupole orderings of $O_2^0$, $O_2^2$ with $\Gamma_3$ or $O_{yz}$, $O_{zx}$, $O_{xy}$ 
with $\Gamma_5$ are interesting phenomena in the $\Gamma_8$ system. 
We refer to CeAg exhibiting the ferroquadrupole (FQ) ordering of $O_2^0$ at $T_{\rm Q}=15$ K.\cite{ref01,ref02} CeB$_6$ is known as 
the antiferroquadrupole (AFQ) ordering of $O_{xy}$-type with the propagation vector of \mbox{\boldmath $k$}=[111] 
at $T_{\rm Q}=3.2$ K.\cite{ref03,ref04} The octupole moments $T_{xyz}$ with $\Gamma_2$ symmetry, $T_x^\alpha$, $T_y^\alpha$, $T_z^\alpha$ 
with $\Gamma_4$ and $T_x^\beta$, $T_y^\beta$, $T_z^\beta$ with $\Gamma_5$ may play a role in the $\Gamma_8$ system.\cite{ref05}

\begin{figure}[t]
\includegraphics[bb=0 0 652 636,width=0.9\linewidth]{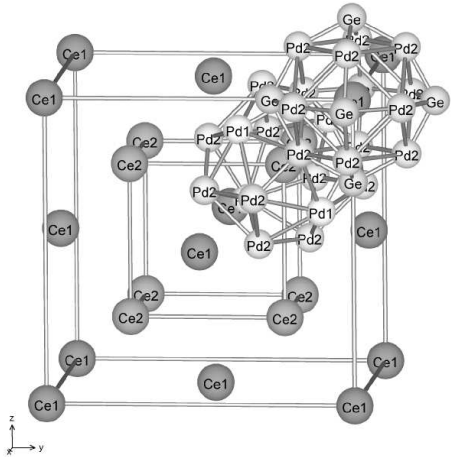}
\caption{\label{eps01} Cage of 4a site with O$\rm_h$ symmetry consisting of Ge and Pd2 atoms and 
cage of 8c site with T$\rm_d$ of Pd1 and Pd2 atoms in clathrate compound Ce$_3$Pd$_{20}$Ge$_6$. 
The 4a site Ce2 forms a simple cubic lattice, while 8c site Ce1 makes a face centered cubic one. 
The ferroquadrupole ordering below $T_{\rm Q1}$ is relevant for the 8c sites and the antiferromagnetic ordering 
below $T_{\rm N2}$ occurs at 4a sites. The $\Gamma_5$ rattling motion originates from the off-center Ce1 atom in 4a-site cage.}
\end{figure}

A cerium-based ternary compound Ce$_3$Pd$_{20}$Ge$_6$ with the $\Gamma_8$ ground state has received much attention 
because the competition between quadrupole and magnetic orderings is expected at low temperatures.\cite{ref06} 
Ce$_3$Pd$_{20}$Ge$_6$ crystallizes in a cubic Cr$_{23}$C$_6$-type structure with a space group $Fm\bar{3}m$ 
consisting of four molecular units with 116 atoms in a unit cell.\cite{ref07} 
The twelve Ce sites located in cages are divided into two nonequivalent sites in crystallography. 
As shown in Fig.~\ref{eps01} the Ce ion at 4a site in a cage consisting of twelve Pd-atoms and six Ge atoms 
possesses point group symmetry O$\rm_h$, while the Ce ion at 8c site in a cage of sixteen Pd atoms has T$\rm_d$. 
The 4a sites form a face-centered cubic lattice, while the 8c sites make a simple cubic lattice. 
Inelastic neutron scattering on Ce$_3$Pd$_{20}$Ge$_6$ revealed overlapping two peaks 
for the crystalline electric field (CEF) potentials, which correspond to magnetic dipole transitions 
from the $\Gamma_8$ ground quartet to the $\Gamma_7$ excited doublet at 60 K of the 4a site and 
from the $\Gamma_8$ ground quartet to the $\Gamma_7$ at 46 K of 8c site.\cite{ref08} 
The entropy obtained by low-temperature specific heat measurement on Ce$_3$Pd$_{20}$Ge$_6$ 
also indicates the ground state $\Gamma_8$ quartet at both 4a and 8c sites.\cite{ref09}

The low-temperature specific heat of Ce$_3$Pd$_{20}$Ge$_6$ shows a rounded small peak at $T_{\rm Q1}=1.25$ K 
and a sharp $\lambda$-peak at $T_{\rm N2}=0.75$ K.\cite{ref09} Magnetic susceptibility shows a clear cusp at $T_{\rm N2}$, 
but exhibits no sign of anomaly at $T_{\rm Q1}$.\cite{ref06} In addition to these experimental results, 
an elastic softening of $(C_{11}-C_{12})/2$ in our preliminary paper suggests that the paramagnetic phase I 
transforms to the FQ phase II at $T_{\rm Q1}$ and successively changes to the antiferromagnetic (AFM) phase III 
at $T_{\rm N2}$.\cite{ref10} The neutron scattering on Ce$_3$Pd$_{20}$Ge$_6$ reveals a paramagnetic state of Ce ions 
at both 4a and 8c sites even in phase II between $T_{\rm Q1}$ and $T_{\rm N2}$. The AFM ordering in phase III 
with a propagation vector $\mbox{\boldmath $k$}_1=[001]$ for cerium ions at 4a site is observed below $T_{\rm N2}$.\cite{ref11} 
Even in phase III below $T_{\rm N2}$, the 8c site still remains to be the paramagnetic state. 
The AFM ordering with incommensurate structure at 8c site appears only below $T_{\rm N}^*=0.45$ K.

The clathrate compounds exhibiting the rattling motion or off-center motion in a cage have received attention 
because their remarkable reduction of thermal conductivity is favorable for application to thermoelectric device 
with a high figure of merit.\cite{ref12} The ultrasonic waves are scattered by the rattling motion in an over-sized cage 
of a semiconductor Sr$_8$Ga$_{16}$Ge$_{30}$ and a filled skutterudite compound PrOs$_4$Sb$_{12}$.\cite{ref13,ref14} 
The off-center tunneling motion of OH ion doped in NaCl gives rise to elastic softening at low temperatures.\cite{ref15} 
The rattling motion in the present compound Ce$_3$Pd$_{20}$Ge$_6$ with clathrate structure has not been reported so far.

In the present paper we show ultrasonic measurements on Ce$_3$Pd$_{20}$Ge$_6$ in order to examine lattice effects 
associated with the quadrupole ordering and rattling motion in the system. The thermal expansion measurement is also 
employed to detect the spontaneous distortion below $T_{\rm Q1}$. In Sec. II, the experimental procedure and apparatus 
are described. The results of the elastic constant, magnetic phase diagram, thermal expansion are presented in Sec. III. 
The ultrasonic dispersion due to rattling motion is also argued in Sec. III. In Sec. IV, we present concluding remarks.

\section{EXPERIMENT}
Single crystals of Ce$_3$Pd$_{20}$Ge$_6$ used in the present measurements were grown by a Czochralski puling method. 
We have made the ultrasonic velocity measurements using an apparatus consisting of a phase difference detector. 
Piezoelectric plates of LiNbO$_3$ for the ultrasonic wave generation and detection are bonded on plane parallel surfaces 
of sample. The {\it x}-cut plate of LiNbO$_3$ is available for transverse ultrasonic waves and the 36$^\circ${\it y}-cut plate is for 
longitudinal waves. The ultrasonic velocity {\it v} was measured by fundamental frequencies of 10 MHz and overtone excitations 
of 30, 50 and 70 MHz. In the estimation of the elastic constant $C=\rho v^2$, we use the mass density $\rho=10.254$ g/cm$^3$ 
for Ce$_3$Pd$_{20}$Ge$_6$ with a lattice parameter $a=12.457$ {\AA}.\cite{ref06}

A homemade $^3$He-refrigerator equipped with a superconducting magnet was used for low-temperature measurements down to 
450 mK in magnetic fields up to 12 T. A $^3$He-$^4$He dilution refrigerator with a top-loading probe was used for 
the ultrasonic measurements in low-temperature region down to 20 mK in fields up to 16 T. 
Low input-power condition provides the low-temperature ultrasonic measurements 
free from a self-heating effect in the ultrasonic transducers. The sample length as a function of temperature 
or applied magnetic field was measured precisely by a capacitance dilatometer in the $^3$He-refrigerator.

\section{RESULTS AND DISCUSSIONS}
\subsection{Temperature dependence \\of the elastic constants}
\begin{figure}
\includegraphics[width=0.9\linewidth]{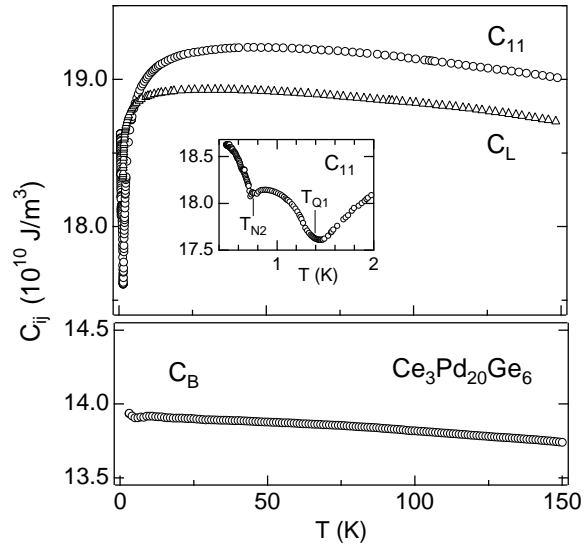}
\caption{\label{eps02} Temperature dependence of the elastic constants $C_{11}$, $C_{\rm L}$ and 
the bulk modulus $C_{\rm B}$ of Ce$_3$Pd$_{20}$Ge$_6$. Inset shows the anomalies of $C_{11}$ 
around the ferroquadrupole transition $T_{\rm Q1}=1.25$ K and the antiferromagnetic transition $T_{\rm N2}=0.75$ K.}
\end{figure}

The elastic constants of $C_{11}$ and $C_{\rm L}=(C_{11}+C_{12}+2C_{44})/2$ of Ce$_3$Pd$_{20}$Ge$_6$ in Fig.~\ref{eps02} 
were measured by the longitudinal ultrasonic waves with frequencies 10 MHz propagating along the [100] and [110] directions, respectively. 
The elastic constant $(C_{11}-C_{12})/2$ of Ce$_3$Pd$_{20}$Ge$_6$ in Fig.~\ref{eps03} was measured by the transverse ultrasonic wave 
of 10 MHz propagating along the [110] direction polarized to the $[1\bar{1}0]$ one. The elastic constant $C_{44}$ of 
Ce$_3$Pd$_{20}$Ge$_6$ in Fig.~\ref{eps03} was determined by the transverse wave of 30 MHz propagating along [100] polarized to [010]. 
The bulk modulus $C_B= (C_{11}+2C_{12})/3$ in Fig.~\ref{eps02} was calculated by $C_{11}$ in Fig.~\ref{eps02} and $(C_{11}-C_{12})/2$ in Fig.~\ref{eps03}.

It is remarkable that $(C_{11}-C_{12})/2$ exhibits a huge softening of 50\% with decreasing temperature down to $T_{\rm Q1}=1.25$ K. 
In phase II below $T_{\rm Q1}$ the ultrasonic echo signal of the $(C_{11}-C_{12})/2$ mode completely disappears 
due to a marked ultrasonic attenuation. The softening of the longitudinal $C_{11}$ and $C_{\rm L}$ modes in Fig. 2 originates from 
the softening of $(C_{11}-C_{12})/2$, because $C_{11}$ and $C_{\rm L}$ involve $(C_{11}-C_{12})/2$ in part. The softening of 
$(C_{11}-C_{12})/2$ above $T_{\rm Q1}$ and the spontaneous tetragonal distortion below $T_{\rm Q1}$, that will be shown 
in Sec. III D, provide evidence for the FQ ordering in phase II. The $C_{44}$ in Fig.~\ref{eps03} also exhibits a softening of 2.5\% 
down to $T_{\rm Q1}$. The low-temperature behavior of $C_{11}$ and $C_{44}$ shown in insets of Figs.~\ref{eps02} and \ref{eps03} 
indicates the transition to the FQ phase II at $T_{\rm Q1}$ and successive transition to the AFM phase III at $T_{\rm N2}=0.75$ K.
On the other hand, $C_{\rm B}$ shows monotonic increase with decreasing temperature.

Neutron scattering on Ce$_3$Pd$_{20}$Ge$_6$ revealed the para-magnetic state for Ce ions at both 4a and 8c sites in phase II, 
which is consistent with the present scenario of the FQ ordering at 8c site in phase II below $T_{\rm Q1}$.\cite{ref11} 
The AFM ordering in phase III at 4a site below $T_{\rm N2}$ has been detected by the neutron scattering. 
It has been proposed that the inter-site quadrupole interaction among 8c sites brings 
about the FQ ordering at 8c sites in phase II and Ce ions at 4a sites still remain to be the 
para-state even in phase II. The inter-site magnetic interaction among 4a sites 
gives rise to the AFM ordering in phase III below $T_{\rm N2}$. The magnetic ordering at 8c sites appears only below 0.4 K. 
We discuss about this transition in the following Sec. III C.

\subsection{Quadrupole susceptibility}
\begin{figure}
\includegraphics[width=0.9\linewidth]{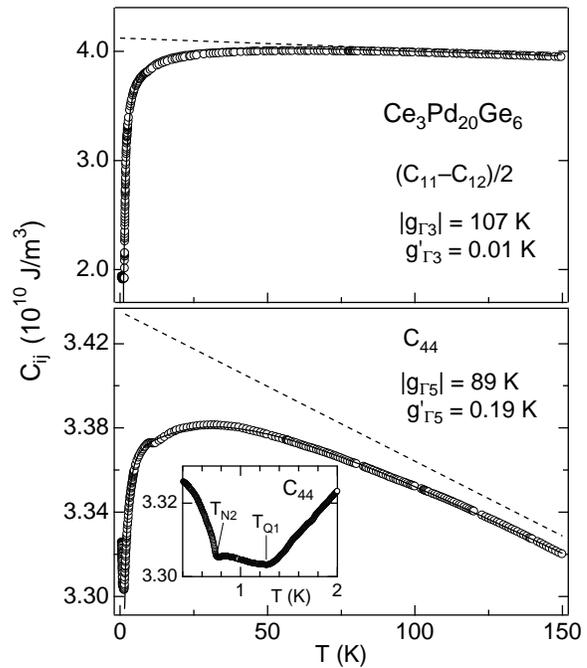}
\caption{\label{eps03} Temperature dependence of the elastic constants $(C_{11}-C_{12})/2$ and $C_{44}$ 
corresponding to shear waves in Ce$_3$Pd$_{20}$Ge$_6$. Inset shows the anomalies of $C_{44}$ 
around the ferroquadrupole transition $T_{\rm Q1}=1.25$ K and the antiferromagnetic transition $T_{\rm N2}=0.75$ K. 
Solid lines are the calculation by the quadrupole susceptibility for the $\Gamma_8$ ground state and $\Gamma_7$ state 
at 46 K of Ce ions. The broken lines were back ground $C_\Gamma ^0$ as shown in the text. A shoulder in $C_{44}$ around 30 K means the ultrasonic dispersion 
due to the $\Gamma_5$ rattling motion.}
\end{figure}

In order to analyze the elastic softening of $(C_{11}-C_{12})/2$ and $C_{44}$ in Ce$_3$Pd$_{20}$Ge$_6$ of Fig.~\ref{eps03}, 
we introduce the coupling of the quadrupole $O_{\Gamma\gamma}$ of Ce ions to the elastic strain 
$\varepsilon_{\Gamma\gamma}$ as ~\cite{ref16}
\begin{eqnarray}
H_{QS}&=&-\sum_i g_\Gamma O_{\Gamma\gamma}(i)\varepsilon_{\Gamma\gamma},
\end{eqnarray}
where the summation $\sum_i$ takes over Ce ions in unit volume and $g_\Gamma$ is a coupling constant. 
The inter-site quadrupole interaction mediated by phonons and conduction electrons 
is written in a mean field approximation as
\begin{eqnarray}
H_{QQ}&=&-\sum_j g'_\Gamma \langle O_{\Gamma\gamma} \rangle O_{\Gamma\gamma}(j),
\end{eqnarray}
where $\langle O_{\Gamma\gamma} \rangle$ denotes a mean field of the quadrupole and $g'_\Gamma$ means a coupling constant for 
the inter-site quadrupole interaction. By differentiating the total free energy consisting of 4f-electron and lattice parts 
with respect to the elastic strain $\varepsilon_{\Gamma\gamma}$, we obtain the temperature dependence of 
the elastic constant $C_\Gamma(T)$ as ~\cite{ref16}
\begin{eqnarray}
C_\Gamma (T)&=&C_\Gamma ^0 -\frac{Ng_\Gamma ^2\chi _\Gamma (T)}{1-g'_\Gamma \chi _\Gamma (T)}.
\end{eqnarray}
Here $C_\Gamma^0(T)$ denotes a background elastic constant without the quadrupole-strain interaction 
and {\it N} is the number of Ce ions in unit volume. The quadrupole susceptibility of $\chi_\Gamma(T)$ in Eq. (3) is  written as
\begin{eqnarray}
-g_\Gamma^2 \chi_\Gamma (T)&=&\langle\frac{\partial^2 E_i}{\partial \varepsilon _{\Gamma\gamma} ^2}\rangle
-\frac{1}{k_BT}\{\langle (\frac{\partial E_i}{\partial \varepsilon_{\Gamma\gamma}})^2\rangle-\langle \frac{\partial E_i}{\partial \varepsilon _{\Gamma\gamma}}\rangle^2\},
\end{eqnarray}
where $E_i$ is a second-order perturbation energy with respect to $\varepsilon_{\Gamma\gamma}$ for CEF state. \cite{ref16} 
The first part in right hand of Eq. (4) corresponds to the Van Vleck-term and the second part to the Curie term. 
The Ce ions at both 4a and 8c sites in Ce$_3$Pd$_{20}$Ge$_6$ have the $\Gamma_8$ ground state, 
while the $\Gamma_7$ state has excited energies of 46 K at 8c site and 60 K at the 4a site. 
As was already mentioned, the neutron scattering revealed that the FQ ordering of 8c sites occurs 
at $T_{\rm Q1}=1.25$ K and the AFM ordering of 4a sites appears at $T_{\rm N2}=0.75$ K. 
These facts indicate that the inter-site quadrupole interaction of Eq. (2) among the Ce ions at 8c sites dominates 
the softening of $(C_{11}-C_{12})/2$ as a precursor of the FQ ordering at $T_{\rm Q1}$. In the following analysis 
we simply assume the quadrupole-strain interaction of Eq. (1) and quadrupole interaction of Eq. (2) 
for the 8c site with the CEF splitting of $\Gamma_8$ (0 K) and $\Gamma_7$ (46 K).

The solid lines for $(C_{11}-C_{12})/2$ and $C_{44}$ with Eq. (3) in Fig.~\ref{eps03} reproduce well the softening 
in paramagnetic phase I above $T_{\rm Q1}$. It should be noted that the softening above $T_{\rm Q1}$ 
proportional to the reciprocal temperature $1/T$ originates from the Curie term of Eq. (4). 
The coupling constants were determined to be $|g_{\Gamma 3}|=107$ K, $g'_{\Gamma 3}=0.01$ K for $(C_{11}-C_{12})/2$ 
and $|g_{\Gamma 5}|=89$ K, $g'_{\Gamma 5}=0.19$ K for $C_{44}$. 
The back ground $(C_{11}^0-C_{12}^0)/2=(4.12-0.001T)\times10^{10}$ J/m$^3$ and 
$C_{44}^0=(3.44-0.0007T)\times10^{10}$ J/m$^3$ indicated by broken lines in Fig.~\ref{eps03} was used.
The positive value of $g'_{\Gamma 3}>0$ are consistent with the FQ ordering in Ce$_3$Pd$_{20}$Ge$_6$. 
A shoulder like anomaly in $C_{44}$ around 10 K results from ultrasonic dispersion that is caused by a rattling motion 
of the rare-earth ion at 4a site in an oversized cage of Fig.~\ref{eps01}. We discuss about this remarkable behavior in Sec. III E. 

\subsection{Magnetic phase diagram}
\begin{figure}
\includegraphics[width=0.9\linewidth]{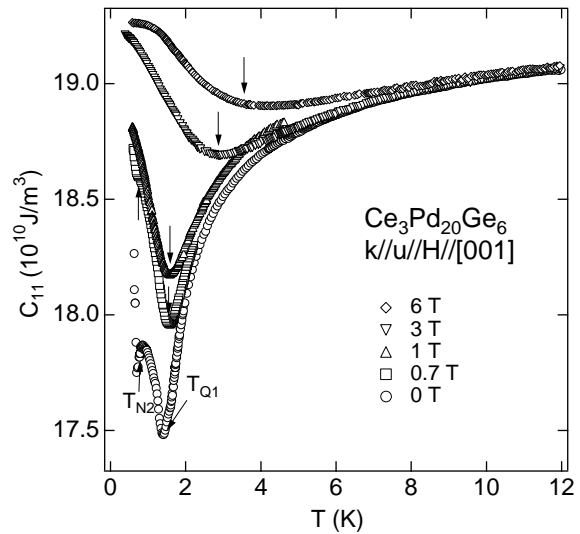}
\caption{\label{eps04} Low-temperature behavior of $C_{11}$ of Ce$_3$Pd$_{20}$Ge$_6$ under magnetic fields 
along the [001] direction. Downward arrows indicate the ferroquadrupole ordering temperature $T_{\rm Q1}$ 
and upward arrow means the antiferromagnetic ordering temperature $T_{\rm N2}$.}
\end{figure}

\begin{figure}
\includegraphics[width=0.9\linewidth]{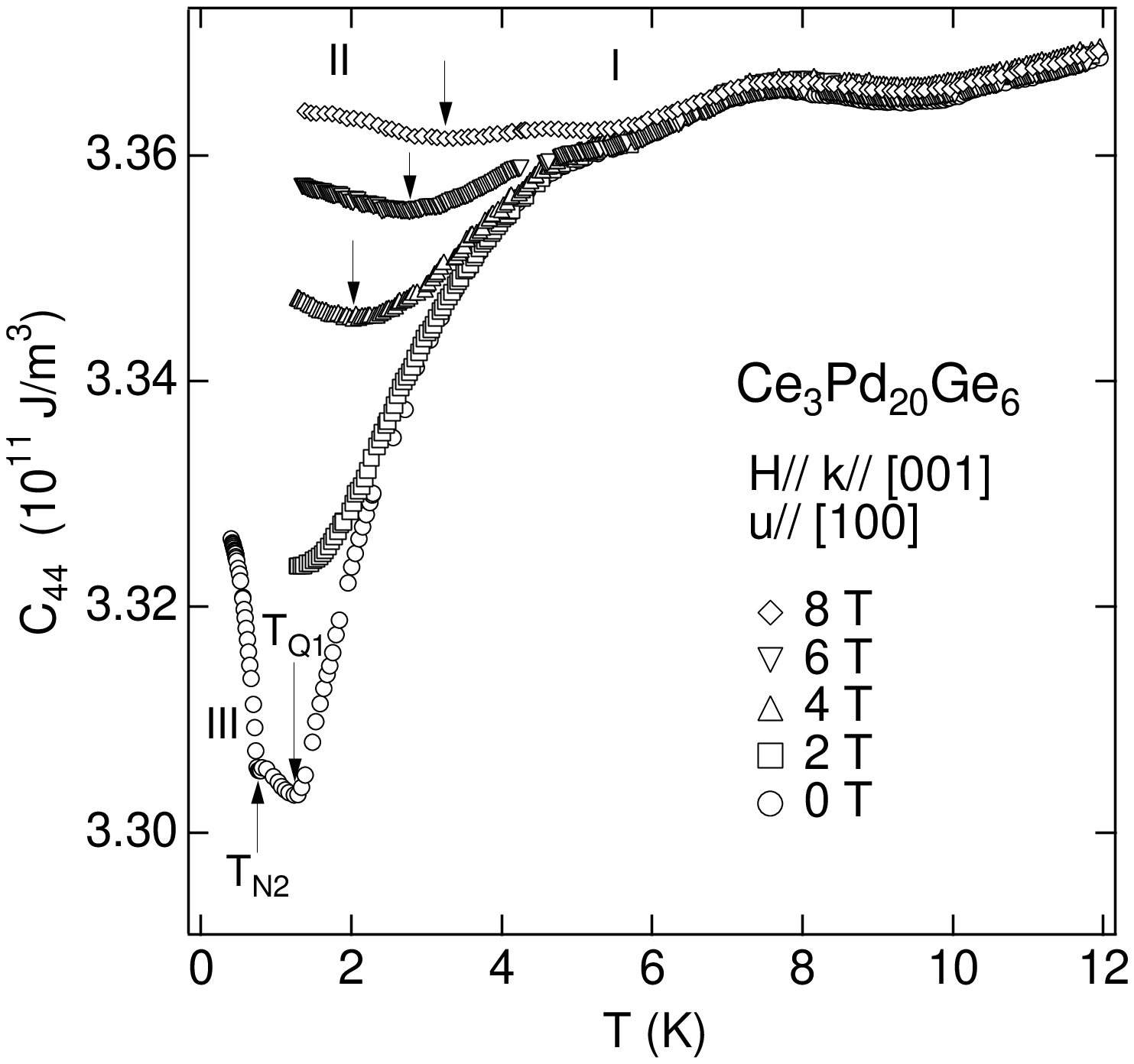}
\caption{\label{eps05} Low-temperature behavior of $C_{44}$ of Ce$_3$Pd$_{20}$Ge$_6$ under magnetic fields along [001]. 
Downward arrows indicate the ferroquadrupole ordering temperature $T_{\rm Q1}$ and upward arrow means 
the antiferromagnetic ordering temperature $T_{\rm N2}$.}
\end{figure}

In order to examine the magnetic phase diagrams of the FQ and AFM orderings in Ce$_3$Pd$_{20}$Ge$_6$, 
we have made the low-temperature ultrasonic measurements of $C_{11}$, $C_{44}$ and $(C_{11}-C_{12})/2$ 
under magnetic fields. The softening of $C_{11}$ of Fig.~\ref{eps04} reduces with increasing fields applied 
along the [001] direction parallel to the propagation direction of longitudinal wave. 
The FQ transition points $T_{\rm Q1}$ indicated by downward arrows in Fig.~\ref{eps04} shift to higher temperatures 
and become indistinct in high fields up to 6 T. In Fig.~\ref{eps05}, the FQ transition points $T_{\rm Q1}$ also shift to 
higher temperatures accompanied by reduction of the softening in $C_{44}$ with increasing applied fields 
along the [001] direction.

\begin{figure}
\includegraphics[width=0.9\linewidth]{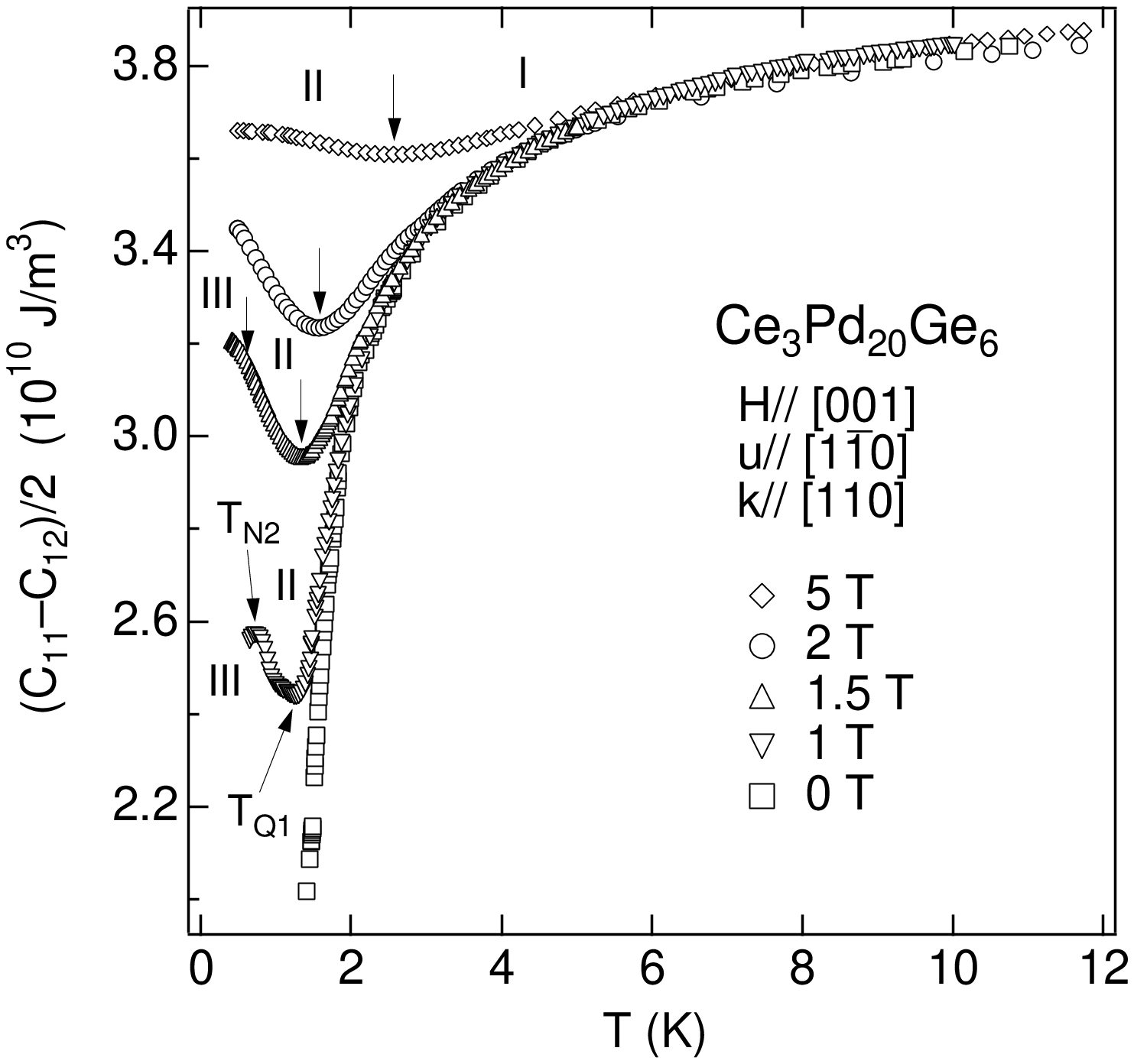}
\caption{\label{eps06} Low temperature behavior of $(C_{11}-C_{12})/2$ of Ce$_3$Pd$_{20}$Ge$_6$ under magnetic fields 
along the [001] direction. The successive phase transitions I-II-III are indicated by arrows.}
\end{figure}

\begin{figure}
\includegraphics[width=0.9\linewidth]{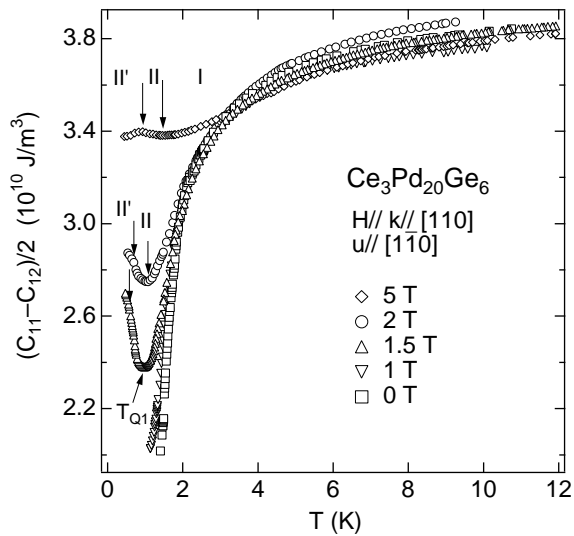}
\caption{\label{eps07} Low temperature behavior of $(C_{11}-C_{12})/2$ of Ce$_3$Pd$_{20}$Ge$_6$ 
under magnetic fields along the [110] direction. The successive phase transitions I-II-II$'$ are indicated by arrows.}
\end{figure}

In Figs.~\ref{eps06} and \ref{eps07}, we show the field dependence of $(C_{11}-C_{12})/2$ applying fields 
along the [001] and [110] directions, respectively. 
In zero magnetic field, the $(C_{11}-C_{12})/2$ mode exhibits the considerable softening of 50\% and 
the strong ultrasonic attenuation losing the echo signal in the vicinity of the FQ transition $T_{\rm Q1}=1.25$ K. 
The magnitude of the softening decreases abruptly with increasing fields along both [001] and [110] directions. 
In magnetic fields, clear minima of $(C_{11}-C_{12})/2$ corresponding to the transitions $T_{\rm Q1}$ 
from the paramagnetic phase I to the FQ phase II shift to higher temperatures. Only broad round anomalies 
around $T_{\rm Q1}$ have been observed in high fields of 5 T. This behavior of $T_{\rm Q1}$ is similar to the results of 
the FQ ordering accompanied by the soft $C_{44}$ mode in HoB$_6$.\cite{ref17} The Neel temperature $T_{\rm N2}$ in Fig.~\ref{eps06} 
shifts slightly to lower temperatures with increasing magnetic fields. In Fig.~\ref{eps07} the anomalies associated 
with the transition between phase II and II$'$ below $T_{Q1}$ have been found. 

\begin{figure}
\includegraphics[width=0.9\linewidth]{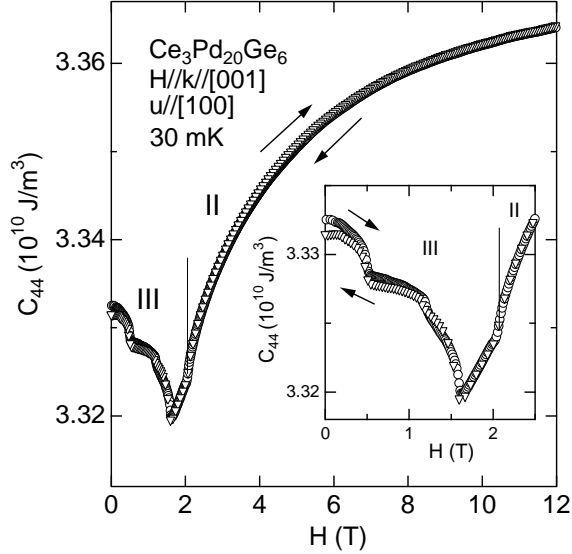}
\caption{\label{eps08} Field dependence of the $C_{44}$ at 30 mK in fields along [001] up to 12 T. 
Inset is expanded view below 2.5 T indicating the magnetic transitions.}
\end{figure}

For the investigation of low-temperature and high-field behavior in FQ II and AFM III phases, 
we have measured field dependence of the $C_{44}$ and $C_{\rm L}=(C_{11}+C_{12}+2C_{44})/2$ 
employing the dilution refrigerator. In Fig.~\ref{eps08} we show $C_{44}$ versus $H$ at 30 mK in fields up to 12 T 
applied along [001]. Inset of Fig. 8 is expanded view below 2.5 T. 
An anomaly of the phase II-III boundary at 2.1 T indicated by a vertical line has been observed. 
Furthermore, several anomalies at 0.5, 1.2 and 1.6 T associated with sub-phases of the phase III have been found. 
It should be emphasized that appreciable hysteresis phenomena between increasing and decreasing field sequences 
have been found only in phase III. 

\begin{figure}
\includegraphics[width=0.8\linewidth]{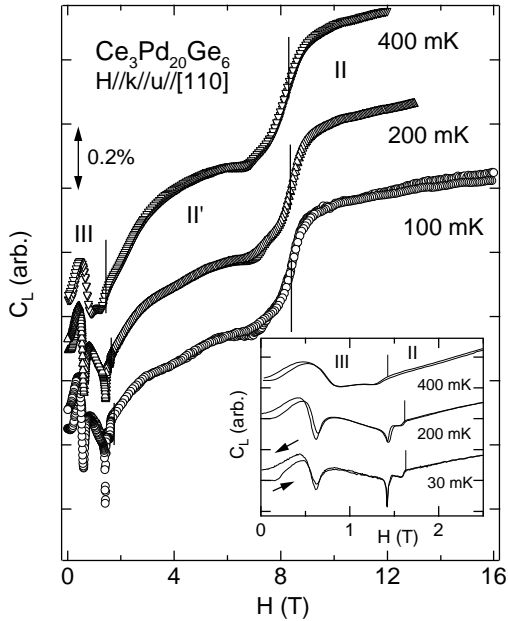}
\caption{\label{eps09} Field dependence of the $C_{\rm L}=(C_{11}+C_{12}+2C_{44})/2$ at various temperatures 
in fields along [110] up to 16 T. Inset is expanded view below 2.5 T indicating the magnetic transitions}
\end{figure}

In Fig.~\ref{eps09} we show the low-temperature field dependence of the $C_{\rm L}$ in fields up to 16 T applied along [110]. 
Low-field behavior below 2.5 T is shown in inset of Fig. 9. 
We have observed a new phase boundary around 8.2 T in phase II, which is probably a sub-phase of the FQ phase II. 
However, this phase boundary is absent in fields along [001] as shown in Fig.~\ref{eps08}. At low field in phase III, 
we have found several anomalies in $C_{\rm L}$ of Fig.~\ref{eps09} showing a hysteresis behavior. 
As can been seen in inset of Fig.~\ref{eps09}, this hysteresis becomes pronounced with decreasing temperature. 
These sub-phases with hysteresis behavior in phase III in magnetic fields along both [001] and [110] are well consistent 
with the results of neutron scattering experiments that detect weak incommensurate magnetic Bragg peaks 
with a propagation vector \mbox{\boldmath $k$}=[0 0 1-$\tau$], ($\tau\sim$ 0.06) at 8c site.\cite{ref11}

\begin{figure}
\includegraphics[width=0.9\linewidth]{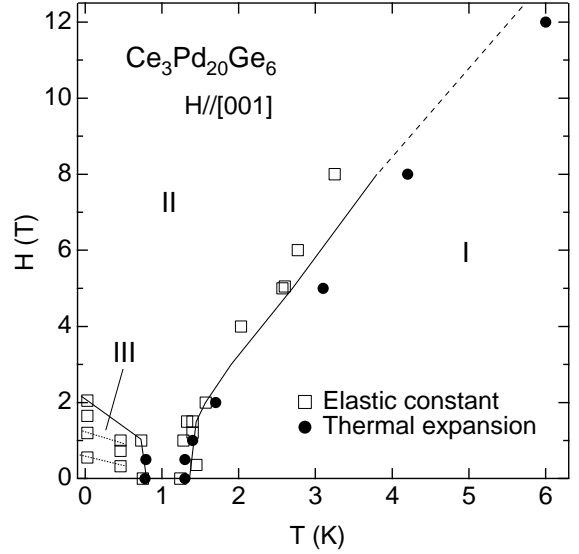}
\caption{\label{eps10} Magnetic phase diagram of Ce$_3$Pd$_{20}$Ge$_6$ under fields along the [001] direction. 
The boundary from paramagnetic phase I to the ferroquadrupole phase II shifts to higher temperatures with increasing fields, 
while the boundary from phase II to the antiferromagnetic phase III shifts to lower temperatures in fields.}
\end{figure}

\begin{figure}
\includegraphics[width=0.9\linewidth]{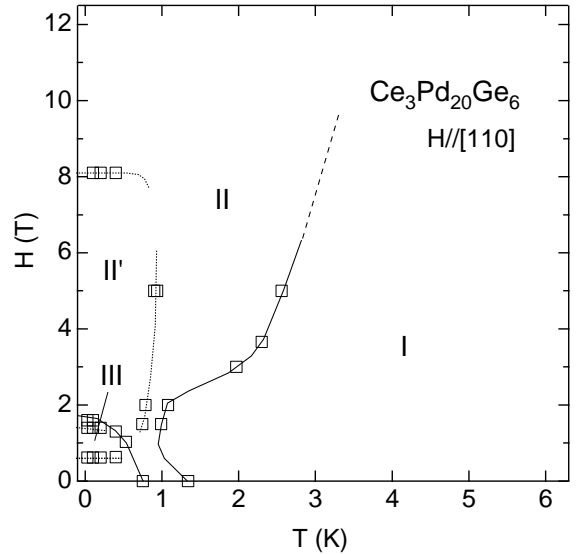}
\caption{\label{eps11} Magnetic phase diagram of Ce$_3$Pd$_{20}$Ge$_6$ under fields along the [110] direction. 
The sub-phase II$'$ exists in the AFQ phase II.}
\end{figure}

The magnetic phase diagrams of Ce$_3$Pd$_{20}$Ge$_6$ in Figs.~\ref{eps10} and \ref{eps11} are obtained in fields 
along the [001] and [110] directions, respectively. We present the results of the ultrasonic measurements 
together with the results of thermal expansions in Sec. III D. It is of importance that the FQ phase II 
is stabilized in fields for the [001] direction in Fig.~\ref{eps10} and the [110] direction in Fig.~\ref{eps11}. 
The FQ sub-phase II$'$ was added to the phase diagram and the upper limit at 8.2 T of the phase II$'$ 
newly determined in fields along [110] of Fig.~\ref{eps11}. 
However, the FQ sub-phase is absent in fields along [001] of Fig.~\ref{eps10}. 
This result indicates strong anisotropy of the quadrupole interaction of $O_2^0$ in Ce$_3$Pd$_{20}$Ge$_6$.

The series of R$_3$Pd$_{20}$X$_6$ compounds usually show two successive AFM orderings of 8c site at $T_{\rm N1}$ 
with a propagation vector $\mbox{\boldmath $k$}_1=[111]$ and of 4a site at $T_{\rm N2} (<T_{\rm N1})$ 
with $\mbox{\boldmath $k$}_2=[001]$, Nd$_3$Pd$_{20}$Ge$_6$ ($T_{\rm N1}=1.75$ K, $T_{\rm N2}=0.58$ K) \cite{ref18}, 
Nd$_3$Pd$_{20}$Si$_6$ ($T_{\rm N1}=2.4$ K, $T_{\rm N2}=0.7$ K) \cite{ref19}, 
Tb$_3$Pd$_{20}$Si$_6$ ($T_{\rm N1}=10.2$ K, $T_{\rm N2}=4.1$ K) \cite{ref20}, 
Dy$_3$Pd$_{20}$Si$_6$ ($T_{\rm N1} \sim 5.8$ K, $T_{N2} \sim 1.8$ K) \cite{ref21} and so on. 
One can reasonably expect that the transition temperature $T_{\rm N1}$ at 8c site is always higher than $T_{\rm N2}$ 
at 4a site since the distance $\sim$ 6.2 {\AA} between rare-earth ion of 8c site is much shorter than 
the one $\sim$ 8.8 {\AA} of 4a site. In the present Ce$_3$Pd$_{20}$Ge$_6$, at first the FQ ordering at 8c site 
with a structural change from cubic lattice to tetragonal one occurs at $T_{\rm Q1}=1.25$ K. 
Therefore, the AFM ordering at 8c site is hard to take place 
because the favorable propagation vector $\mbox{\boldmath $k$}_1=[111]$ 
of 8c site does not match to the tetragonal lattice in phase II. In other words, the AFM ordering at 8c site is 
replaced by the FQ ordering in Ce$_3$Pd$_{20}$Ge$_6$. While, the AFM ordering of 4a site 
with a propagation vector $\mbox{\boldmath $k$}_2=[001]$ is easy to occur even in tetragonal structure below $T_{\rm N2}$. 
Neutron experiments detected large enough value of saturation cerium moments $\mu(4a)=(1.1\pm0.1)\mu_{\rm B}$/Ce 
that is expected from the ground state quartet $\Gamma_8$ perpendicular to the $\mbox{\boldmath $k$}_1=[001]$ 
in Ce$_3$Pd$_{20}$Ge$_6$ far below $T_{\rm N2}$ at 50 mK.\cite{ref11}

\subsection{Thermal expansion}
\begin{figure}[b]
\includegraphics[width=0.9\linewidth]{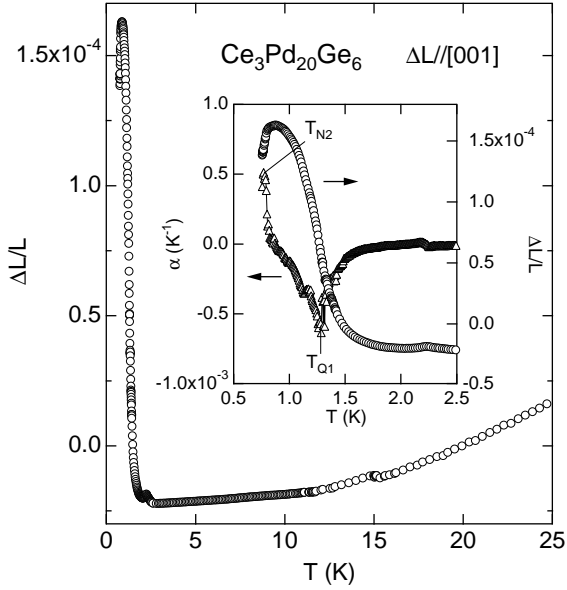}
\caption{\label{eps12} Thermal expansion $\Delta L/L$ in Ce$_3$Pd$_{20}$Ge$_6$. 
Inset shows the thermal expansion coefficient $\alpha$ and thermal expansion $\Delta L/L$ at low temperatures.}
\end{figure}

In order to examine the structural change due to the FQ ordering at $T_{\rm Q1}=1.25$ K, 
we have measured the thermal expansion along the [001] direction in Ce$_3$Pd$_{20}$Ge$_6$. 
The sample lengths along the [001] and [111] directions are written by the symmetry strains as 
$(\Delta L/L)_{[001]}=\varepsilon_{zz}=\varepsilon_{\rm B}/3+\varepsilon_u/\sqrt3$ and 
$(\Delta L/L)_{[111]}=\varepsilon_{\rm B}/3+2(\varepsilon_{yz}+\varepsilon_{zx}+\varepsilon_{xy})/3$. 
Here, $\varepsilon_{\rm B} = \varepsilon_{xx} + \varepsilon_{yy} + \varepsilon_{zz}$ is a volume strain with $\Gamma_1$ symmetry, 
$\varepsilon_u =(2\varepsilon_{zz} - \varepsilon_{xx} - \varepsilon_{yy})/\sqrt3$ is a tetragonal strain with $\Gamma_3$ and 
$\varepsilon_{xy}$ is a shear strain with $\Gamma_5$. The length along [001] in Fig.~\ref{eps12} shows a monotonous decrease with 
decreasing temperature in paramagnetic phase I above $T_{\rm Q1}$ and abruptly expands 
about $\Delta L/L=1.9\times10^{-4}$ below $T_{\rm Q1}$. The thermal expansion along [001] in phase II 
and the huge softening of 50\% in $(C_{11}-C_{12})/2$ of Fig.~\ref{eps03} indicate the $O_2^0$-type FQ ordering accompanied 
by the structural transition from cubic lattice to tetragonal one with the spontaneous strain 
$\langle\varepsilon_u\rangle$ in phase II. This spontaneous strain is proportional to the order parameter as 
$\langle\varepsilon_u\rangle=Ng_{\Gamma3} \langle O_2^0 \rangle (2/(C_{11}^0-C_{12}^0))$ in mean-field approximation. 
Below $T_{\rm N2}=0.75$ K, the $\Delta L/L$ along [001] slightly shrinks. Inset of Fig.~\ref{eps12} is expanded view of $\Delta L/L$ 
and the coefficient of the thermal expansion $\alpha$ at low temperatures. 
A sharp anomaly in the coefficient $\alpha$ has been found at the FQ transition $T_{Q1}$.

\begin{figure}[b]
\includegraphics[width=0.9\linewidth]{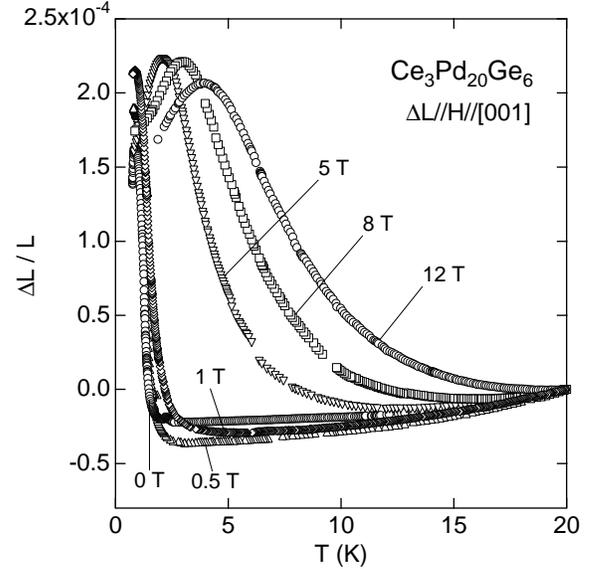}
\caption{\label{eps13} Thermal expansion $\Delta L/L$ along [001] in Ce$_3$Pd$_{20}$Ge$_6$ under fields up to 12 T 
along the [001] direction .}
\end{figure}

Measurements of $\Delta L/L$ versus $T$ in various magnetic fields parallel to [001] are shown in Fig.~\ref{eps13}. 
The magnitude of the expansion $\Delta L/L$ in fields exhibits noticeable increase up to 
$\Delta L/L=2.5\times10^{-4}$ compared with that in zero magnetic field. The sharp increase of $\Delta L/L$ 
at the transition point to the FQ phase II has been observed in low magnetic fields below 1 T. 
On the other hand, the gradual increase in the thermal expansion $\Delta L/L$ above 5 T up to 12 T indicates 
an obscure character of the I-II phase boundary in high fields. 
This is consistent with the fact that the elastic constants in fields of Figs.~\ref{eps04}, \ref{eps05}, \ref{eps06} 
and \ref{eps07} show obscure transitions in fields. 
This behavior is similar to the liquid-gas transition near the critical end point under hydrostatic pressures. 

The thermal expansion along the [001] direction in fields parallel to [001] and the considerable softening of 
$(C_{11}-C_{12})/2$ of 50\% in Ce$_3$Pd$_{20}$Ge$_6$ strongly suggests that the order parameter of the FQ ordering 
in phase II is $O_2^0$ with $\Gamma_3$ symmetry. The relatively small softening of 2.5\% in $C_{44}$ in Fig.~\ref{eps03} 
means that the quadrupole of $O_{xy}$-type with $\Gamma_5$ is irrelevant for the transition at $T_{\rm Q1}$. 
The thermal expansion of Ce$_3$Pd$_{20}$Ge$_6$ along [111] is required to examine an interplay of the spontaneous strain 
$\varepsilon_{xy}$ for the phase II. We refer to our recent study of the FQ transition in HoB$_6$ 
and the phase IV in Ce$_x$La$_{1-x}$B$_6$ ({\it x}=0.75, 0.70), \cite{ref17,ref22} where the trigonal strain 
$\varepsilon_{yz}=\varepsilon_{zx}=\varepsilon_{xy}$ plays a significant role and 
the tetragonal strain $\varepsilon_u$ is irrelevant. These facts are well consistent with 
the pronounced elastic softening in $C_{44}$ of 70\% in HoB$_6$ and of 31\% in Ce$_x$La$_{1-x}$B$_6$ ({\it x}=0.75, 0.70).

\subsection{Ultrasonic dispersion of $\mbox{\boldmath $C_{44}$}$}
\begin{figure}
\includegraphics[width=0.9\linewidth]{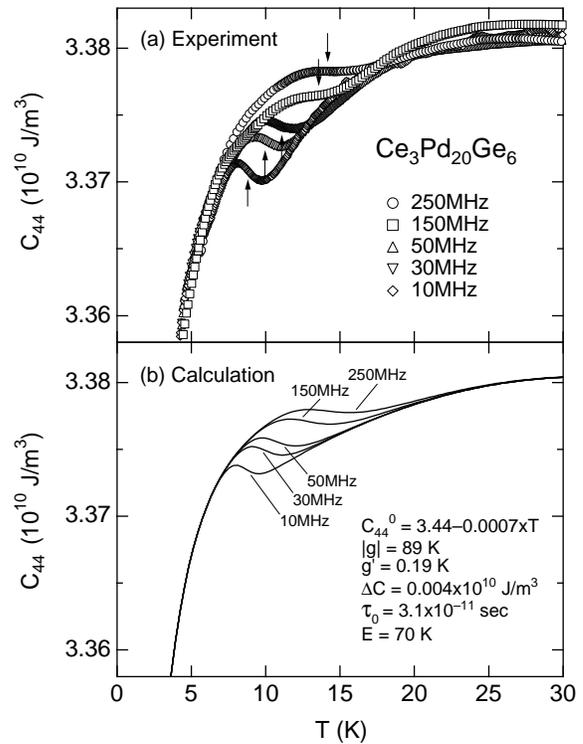}
\caption{\label{eps14} Frequency dependence of the elastic constant of the transverse $C_{44}$ mode 
in Ce$_3$Pd$_{20}$Ge$_6$. (a) represents the experimental results with the frequencies from 10 MHz up to 250 MHz. 
(b) is the calculation in terms of the Debye-type dispersion of Eq. (5) in text.}
\end{figure}

The $C_{44}$ mode associated with the elastic strain $\varepsilon_{yz},\varepsilon_{zx},\varepsilon_{xy}$ 
of $\Gamma_5$ symmetry of Ce$_3$Pd$_{20}$Ge$_6$ in Figs.~\ref{eps03} and \ref{eps05} 
exhibits a shoulder like anomaly around 10 K in addition to the characteristic softening 
due to the quadrupolar coupling above $T_{\rm Q1}=1.25$ K. It should be noted that this anomaly is absent 
for the $(C_{11}-C_{12})/2$ mode associated with $\Gamma_3$ elastic strain $\varepsilon_v$ and 
the bulk modulus $C_{\rm B}$ with $\Gamma_1$ volume strain $\varepsilon_{\rm B}$. In order to examine the origin of 
this anomaly, we have measured the frequency dependence of $C_{44}$ from 10 MHz up to 250 MHz. 
The elastic constant $C_{44}$ of Fig.~\ref{eps14}(a) exhibits shoulders showing remarkable frequency dependence. 
An increase in ultrasonic attenuation around shoulder has also been found, but not discussed here. 
We describe this frequency dependence of the elastic constant $C_{44}^D(\omega)$ in terms of Debye-type dispersion as
\begin{eqnarray}
C_{44}^D(\omega)&=&C_{44}^D(\infty)-\frac{C_{44}^D(\infty)-C_{44}^D(0)}{1+\omega^2\tau^2},
\end{eqnarray}
where $C_{44}^D(\infty)$ and $C_{44}^D(0)$ are the elastic constants of high frequency limit and low frequency one, respectively. 
Here $\omega$ is an angular frequency of the ultrasonic wave and $\tau$ means the relaxation time of the system. 
In fittings of Fig.~\ref{eps14}(b), we take into account the superposition of two susceptibilities by the quadrupole one 
of Eq.(3) and the Debye-type dispersion of Eq.(5) as $C_{44} = C_{44}^Q + C_{44}^D(\omega)$. 
The inflection points around 10 K in $C_{44}$ indicated by arrows in Fig.~\ref{eps14}(a) mean the temperatures 
where the $\tau$ coincides with the $\omega$ as $\omega\tau=1$. 
The ultrasonic attenuation is expected to be maximum at the temperatures for $\omega\tau=1$. 
The solid lines of Fig.~\ref{eps14}(b) being the calculations with Eq. (5) well reproduce the experimental results of Fig.~\ref{eps14}(a). 
The relaxation time obeying the Arrhenius-type temperature dependence $\tau=\tau_0$exp$(E/k_BT)$ 
with the attempt time $\tau_0=3.1\times10^{-11}$ sec and the activation energy $E=70$ K has been determined. 
The parameter of $\Delta C=C_{44}^D(\infty)-C_{44}^D(0)=0.004\times10^{10}$ J/m$^3$ is used. 

The ultrasonic dispersion due to electron thermal hopping has already found 
in the inhomogeneous valence fluctuation compounds of Sm$_3$X$_4$ (X=Se, Te), 
Yb$_4$(As$_{0.71}$Sb$_{0.29}$)$_3$ \cite{ref23} and Sr$_{12}$Ca$_2$Cu$_{24}$O$_{41}$ \cite{ref24}. 
It is remarkable that the very slow relaxation time $\tau$ and extremely low activation energy $E$ 
for Ce$_3$Pd$_{20}$Ge$_6$ are exceptional as compared to those of charge fluctuation compounds 
Sm$_3$Se$_4$ \cite{ref25} and Sm$_3$Te$_4$ \cite{ref26}, $\tau_0 \sim 2.5\times10^{-13}$ sec and $E \sim 1600$ K, 
and Sr$_{12}$Ca$_2$Cu$_{24}$O$_{41}$, $\tau_0=1.01\times10^{-13}$ sec and $E=1900$ K.\cite{ref24} 
This discrepancy of the order of $\tau_0$ and $E$ between the present Ce$_3$Pd$_{20}$Ge$_6$ 
and the charge fluctuation compounds indicates that thermally activated rattling motion of heavy mass particle, 
which is probably rare-earth ion in a cage, gives rise to the ultrasonic dispersion in Ce$_3$Pd$_{20}$Ge$_6$.

Glass materials and charge fluctuation compounds exhibit frequently the ultrasonic dispersion, 
which results from a thermally activated motion in a double- or multi-well potential. 
The compounds such as Sm$_3$X$_4$ (X=S, Se, Te) with different valence of Sm$^{2+}$ and Sm$^{3+}$ ions and 
Sr$_{12}$Ca$_2$Cu$_{24}$O$_{41}$ with Cu$^{2+}$ and Cu$^{3+}$ ions cause the ultrasonic dispersions 
due to thermally assisted charge fluctuation in the temperature region between 100-200 K. 
The two-level system (TLS) due to an atomic tunneling \cite{ref27} or electron tunneling \cite{ref26} manifests itself 
in glass materials at low temperatures, where the thermally activated motion dies out. 
The TLS yields the decrease in the elastic constant proportional to ln$T$,\cite{ref28} 
the specific heat to $T$ and thermal conductivity to $T^2$.\cite{ref29} Besides in the case of Sm$_3$Te$_4$, 
remarkable logarithmic decrease in the elastic constant appears below about 15 K down to the spin glass transition 
at $T_g=1.5$ K, which suggests the existence of the 4f-electron tunneling motion between Sm$^{2+}$ and Sm$^{3+}$ ions 
situated charge glass state.\cite{ref26} 

\begin{table*}
\caption{\label{tab:table1}
Rattling modes with eight off-center positions along the three-fold [111] direction in the 4a-site cage 
with O$\rm_h$ point group symmetry. Corresponding elastic strains are also listed. 
There is a $\Gamma _5$ rattling mode coupled to the $\varepsilon _{xy}$-type strain of the $C_{44}$ mode. 
This is contrary to the absence of the $\Gamma _3$ rattling mode to the $\varepsilon _u, \varepsilon _v$ 
of the $(C_{11}-C_{12})/2$ mode. The present ultrasonic dispersion in $C_{44}$ of Ce$_3$Pd$_{20}$Ge$_6$ 
originates from the $\Gamma _5$-type rattling at 4a site.}
\begin{ruledtabular}
\begin{tabular}{ccc}
Symmetry & Rattling mode & Strain\\
\hline
$\Gamma _1$ & $\rho _{\Gamma 1}=\rho _1+\rho _2+\rho _3+\rho _4+\rho _5+\rho _6+\rho _7+\rho _8$ & $\varepsilon _{\rm B}=\varepsilon _{xx} + \varepsilon _{yy} + \varepsilon _{zz}$\\
$\Gamma _2$ & $\rho _{\Gamma 2}=\rho _1+\rho _2+\rho _3+\rho _4-\rho _5-\rho _6-\rho _7-\rho _8$ &\\
$\Gamma _3$ &  & $\varepsilon _u=(2\varepsilon _{zz}-\varepsilon _{xx}-\varepsilon _{yy})/\sqrt{3}$\\
&  & $\varepsilon _v=\varepsilon _{xx}-\varepsilon _{yy}$\\
$\Gamma _4$ & $\rho _{\Gamma 4,x}=\rho _1-\rho _2-\rho _3+\rho _4+\rho _5-\rho _6+\rho _7-\rho _8$ & \\
& $\rho _{\Gamma 4,y}=\rho _1-\rho _2+\rho _3-\rho _4+\rho _5-\rho _6-\rho _7+\rho _8$ & \\
& $\rho _{\Gamma _4,z}=\rho _1+\rho _2-\rho _3-\rho _4-\rho _5-\rho _6+\rho _7+\rho _8$ & \\
$\Gamma _5$ & $\rho _{\Gamma _5,yz}=\rho _1-\rho _2-\rho _3+\rho _4-\rho _5+\rho _6-\rho _7+\rho _8$ & $\varepsilon _{yz}$\\
& $\rho _{\Gamma _5,zx}=\rho _1-\rho _2+\rho _3-\rho _4-\rho _5+\rho _6+\rho _7-\rho _8$ & $\varepsilon _{zx}$\\
& $\rho _{\Gamma _5,xy}=\rho _1+\rho _2-\rho _3-\rho _4+\rho _5+\rho _6-\rho _7-\rho _8$ & $\varepsilon _{xy}$\\
\end{tabular}
\end{ruledtabular}
\end{table*}

The present clathrate compound Ce$_3$Pd$_{20}$Ge$_6$ is a crystal possessing an ideal periodic arrangement of cages in space. 
The stable trivalent Ce ions in cages of Ce$_3$Pd$_{20}$Ge$_6$ are free from the valence fluctuation phenomena. 
As shown in Fig.~\ref{eps01}, the clathrate compound Ce$_3$Pd$_{20}$Ge$_6$ is made up of the cage at 4a site consisting of Pd and Ge 
with distances $d\rm_{Ce1-Ge}$=3.332 {\AA}, $d\rm_{Ce1-Pd2}$=3.067 {\AA}, and the cage at 8c site of Pd with $d\rm_{Ce2-Pd1}$=2.868 {\AA}, 
$d\rm_{Ce2-Pd2}$=3.373 {\AA}. The trivalent Ce-ion with radii $a=1.7\sim1.8$ {\AA} inside the 4a-site cage in particular 
is expected to show the rattling motion over the off-center positions being away from the center of the cage. 
Actually the neutron scattering on Pr$_3$Pd$_{20}$Ge$_6$ and Nd$_3$Pd$_{20}$Ge$_6$ revealed the sharp transition peaks 
indicating the stable CEF splitting at 8c site and no indication for CEF state at 4a site.\cite{ref08} 
These results may imply the obscure CEF state due to the off-center Ce ion at 4a site contrary to 
the well-defined CEF splitting at 8c site being stable Ce ion position. 

\subsection{${\bold \Gamma_5}$ rattling motion}
Notable finding of Fig.~\ref{eps14} is that the ultrasonic dispersion in the $C_{44}$ mode associated with $\varepsilon_{xy}$-type 
strain indicates the rattling motion with specific $\Gamma_5$ symmetry in Ce$_3$Pd$_{20}$Ge$_6$. 
It is expected that the Ce ion in 4a cage with cubic symmetry O$\rm_h$ favors off-center positions 
along one of the three principle directions of [100], [110] and [111]. 
As one can see the cage at 4a site in Fig.~\ref{eps01}, it is of particular interest that no atom exists 
along the three-fold [111] directions, while the Ge atom occupies along the four-fold [100] and the Pd atom 
along the two-fold [110] ones. This crystallographic character may promise a flat potential 
along the three-fold [111] directions and profound potentials along the four-fold [100] and two-fold [110] directions. 
Presumably the Ce ion at 4a site prefers the off-center eight positions along the [111] directions, 
which are defined as $r_1=(a,a,a)$, $r_2=(-a,-a,a)$, $r_3=(-a,a,-a)$, $r_4=(a,-a,-a)$, $r_5=(a,a,-a)$, $r_6=(-a,-a,-a)$, 
$r_7=(a,-a,a)$, $r_8=(-a,a,a)$. The atomic densities $\rho_i=\rho_i(r_i) (i=1\sim8)$
at the eight off-center positions are also defined.
When 48 symmetry operators of O$\rm_h$ point group are acted on the atomic densities $\rho_i$, 
one can derive the transfer representation matrices with $8\times8$ elements. Consequently, 
one obtains the characters $\chi_{rat}^{[111]}$ for the rattling motion by tracing the diagonal elements 
of the representation matrices. Using the characters $\chi_{rat}^{[111]}$ and the character table 
for the irreducible representations of O$\rm_h$,\cite{ref30} the rattling motion over the eight off-center positions 
along the [111] direction is reduced to the direct sum of $\Gamma_1$(1D)$\oplus$$\Gamma_2$(1D)$\oplus$$\Gamma_4$(3D)$\oplus$$\Gamma_5$(3D).

Employing projection operators on appropriate atomic density $\rho_i$, we obtain the rattling modes 
for the irreducible representations as listed in Table 1 together with the elastic strains $\varepsilon_{\Gamma\gamma}$. 
One can see the presence of $\Gamma_5$ rattling mode $\rho_{\Gamma 5,yz},\rho_{\Gamma 5,zx},\rho_{\Gamma 5,xy}$ 
coupled to the strain $\varepsilon_{yz}$, $\varepsilon_{zx}$, $\varepsilon_{xy}$ contrary to the absence of $\Gamma_3$ 
rattling mode coupled to the strain $\varepsilon_u$, $\varepsilon_v$. 
This is consistent with the fact that the ultrasonic dispersion reveals in $C_{44}$ and is absent in $(C_{11}-C_{12})$/2. 
The Ce atom in the cage of Ce$_3$Pd$_{20}$Ge$_6$ obeys a harmonic oscillation of $\zeta(z)=(1/\pi z_0)1/2$exp$(-z^2/2z_0^2)$ 
with a mean square displacement $z_0=(1/2\pi)(h\tau_0/M)^{1/2}$. The attempt time $\tau_0=3.1\times10^{-11}$ sec 
and the mass $M=140m_{\rm p}$ where $m_{\rm p}$ is a proton mass, leads to the mean square displacement $z_0$ 
being approximately twice of off-center distances $a$ as $z_0=2a=0.48$ {\AA}.\cite{ref31}

\begin{figure}
\includegraphics[width=0.9\columnwidth]{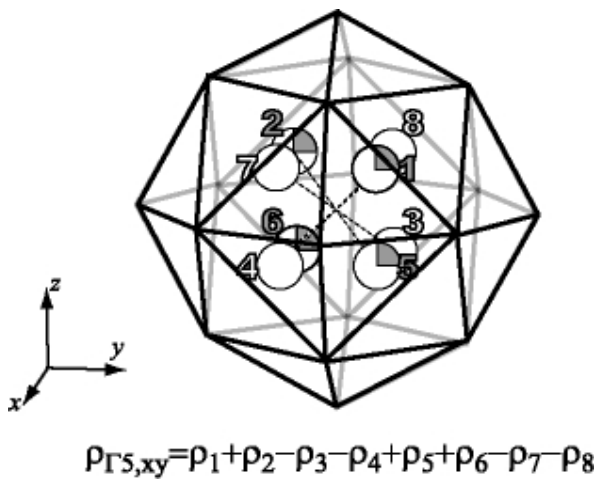}
\caption{\label{rat} Schematic view for the $\Gamma_5$ rattling mode $\rho_{\Gamma 5,xy}$ due to the off-center Ce-ion 
along the three-fold [111] direction in the 4a-site cage. The $\rho_{\Gamma 5,xy}$ represents that fractional atomic density 
1/4 is located at $r_1$, $r_2$, $r_5$, $r_6$ and null at $r_3$, $r_4$, $r_7$, $r_8$. 
The freezing of the thermally activated motion of the $\Gamma_5$ rattling mode brings about the atomic tunneling state at low temperatures.}
\end{figure}

The full symmetry $\Gamma_1$ rattling mode 
$\rho_{\Gamma 1}=\rho_1+\rho_2+\rho_3+\rho_4+\rho_5+\rho_6+\rho_7+\rho_8$ means the uniform atomic distribution 
with fraction 1/8 at each eight off-center positions. While the $\Gamma_5$ rattling mode, for instance 
$\rho_{\Gamma 5,xy}=\rho_1+\rho_2-\rho_3-\rho_4+\rho_5+\rho_6-\rho_7-\rho_8$, represents anisotropic atomic distribution 
being quadrupole $O_{xy}$ at the lowest order such as fraction 1/4 at $r_1$, $r_2$, $r_5$, $r_6$ and zero at $r_3$, $r_4$, $r_7$, $r_8$ 
as shown in Fig.~\ref{rat}. The present group theoretical analysis for the rattling mode is essentially the same treatment 
previously argued for the charge fluctuation mode.\cite{ref24,ref26,ref32} In the present system Ce$_3$Pd$_{20}$Ge$_6$, 
the thermally activated $\Gamma_5$ rattling mode may be a ground state and the $\Gamma_1$, $\Gamma_2$, $\Gamma_4$ 
be excited states. 

Our group has recently found similar ultrasonic dispersion around 30 K in a heavy fermion superconductor PrOs$_4$Sb$_{12}$ 
with a filled skutterudite structure. It should be noted that ultrasonic dispersion in the $(C_{11}-C_{12})/2$ mode of 
PrOs$_4$Sb$_{12}$ indicates the $\Gamma_3$ rattling motion of Pr atom over six fractionally occupied positions along [100]. 
The dispersion of $(C_{11}-C_{12})/2$ in PrOs$_4$Sb$_{12}$ is contrary to the one of $C_{44}$ 
in the present compound of Ce$_3$Pd$_{20}$Ge$_6$.\cite{ref14} The attempted time $\tau_0=8.8\times10^{-11}$ sec, 
activation energy $E=168$ K and mean square displacement $z_0=0.80$ {\AA} in PrOs$_4$Sb$_{12}$ are comparable to 
the present results of Ce$_3$Pd$_{20}$Ge$_6$.

The thermally activated $\Gamma_5$ rattling motion with fractional atomic state in Ce$_3$Pd$_{20}$Ge$_6$ dies out 
with decreasing temperature. At further low temperatures, the off-center tunneling state of Ce ions in the 4a-site cages will appear,
which means a quantum state being occupied four positions, 
for instance at $r_1$, $r_2$, $r_5$, $r_6$ for the case of $\rho_{\Gamma 5,xy}$, at the same time.
The charge glass state in the inhomogeneous mixed valence compound Sm$_3$Te$_4$ revealed 
the low-temperature softening in elastic constants proportional to ln$T$ being resemble the structural glass.\cite{ref26} 
The present scenario of the atomic tunneling of Ce ions in 4a-site cages in Ce$_3$Pd$_{20}$Ge$_6$ is also expected to 
show the elastic softening at low temperatures. However, the low-temperature quadrupole and magnetic orderings in Ce$_3$Pd$_{20}$Ge$_6$ 
may mask the character of the tunneling state in the present case. In order to clarify the tunneling and rattling 
in the present clathrate compounds, the low-temperature thermodynamic and ultrasonic measurements on La$_3$Pd$_{20}$Ge$_6$ 
without 4f-electron in particular is required.

\section{CONCLUDING REMARKS}
In the present paper we have measured the elastic constants and thermal expansion of Ce$_3$Pd$_{20}$Ge$_6$. 
The characteristic elastic softening in $(C_{11}-C_{12})/2$ and $C_{44}$ is well described 
in terms of the quadrupole susceptibility for the $\Gamma_8$ ground state. 
The important finding is that the $(C_{11}-C_{12})/2$ shows the huge softening of 50\% towards $T_{\rm Q1}=1.25$ K 
and the $C_{44}$ exhibits the softening of 2.5\% only. This result strongly indicates the FQ ordering 
with the order parameter of the $\Gamma_3$ symmetry in phase II below $T_{\rm Q1}$. 
Actually we have successfully observed the sharp increase of $\Delta L/L=1.9\times10^{-4}$ 
in length along the [001] direction below $T_{\rm Q1}$. This is the evidence for the $O_2^0$-type FQ ordering 
accompanied by the structural change from cubic lattice to tetragonal one at $T_{\rm Q1}$ in Ce$_3$Pd$_{20}$Ge$_6$. 

For the investigation of the magnetic phase diagram concerning the FQ phase II and AFM phase III in Ce$_3$Pd$_{20}$Ge$_6$, 
the elastic constants and thermal expansion in fields have been measured. 
We have found that the boundary from the paramagnetic phase I to the FQ phase II of $O_2^0$ 
shifts to higher temperatures with increasing magnetic fields. The result that the I-II phase transition 
becomes obscure in fields is similar to the liquid-gas transition approaching to the critical end point under pressure. 
This result consistent with the fact that the FQ order parameter $O_2^0$ in Ce$_3$Pd$_{20}$Ge$_6$ 
has the total symmetry under fields along the [001] direction.\cite{ref05} 
The boundary from the phase II to the AFM ordering shifts to lower temperatures as similar as the conventional AFM ordering.

We have found the ultrasonic dispersion in the $C_{44}$ mode indicating the rattling motion with $\Gamma_5$ symmetry. 
Taking into account the absence of the atom along the [111] direction in 4a-site cage, 
we have successfully picked up the specific $\Gamma_5$ rattling mode 
$\rho_{\Gamma 5,yz}$, $\rho_{\Gamma 5,zx}$, $\rho_{\Gamma 5,xy}$ 
with the fractional atomic density over the eight minimum positions of potential 
along the four-fold [111], $[\bar{1}11]$, $[1\bar{1}1]$ and $[11\bar{1}]$ directions. 
The dispersion of the $C_{44}$ mode obeying the Debye-formula revealed the thermally activated-type relaxation time 
$\tau=\tau_0$exp$(E/k_BT)$ for the $\Gamma_5$ rattling mode with an attempt time $\tau_0=3.1\times10^{-11}$ sec 
and an activation energy $E=70$ K. The estimated mean square displacement $z_0=0.48$ {\AA} for the harmonic oscillation 
of Ce atom leads to the distance of the potential minima along the [111] direction as $a = z_0/2 =0.24$ {\AA}. 
In order to confirm the anisotropic atomic distribution in the 4a-site cage, the neutron or x-ray scattering is required. 
The freezing of the thermally activated motion due to the $\Gamma_5$ rattling mode with lowering temperature 
brings about the atomic tunneling state. 
By analogy of the charge glass compound Sm$_3$Te$_4$, the Ce-ion tunneling is expected at low temperatures. 
The ultrasonic investigation on La$_3$Pd$_{20}$Ge$_6$ free from the long-range ordering due to 4f-electrons is now 
in progress to shed light on the tunneling and rattling in cages.

\begin{acknowledgments}
This work was supported by a Grant-In-Aid for Scientific Research 
from the Ministry of Education, Culture, Sports, Science and Technology of Japan.
\end{acknowledgments}

\end{document}